\documentclass[paper]{revtex4}
\usepackage{amsmath}
\usepackage{amsfonts}
\usepackage{graphicx}

\begin{document}
\tolerance=5000
\def\pp{{\, \mid \hskip -1.5mm =}}
\def\cL{{\cal L}}

\def\nn{\nonumber \\}



\title{Warm tachyonic inflation in warped background}

\author{Atri Deshamukhya$^1$} \author{Sudhakar Panda$^2$}
\affiliation{$^1$ Department of Physics,Assam University,
Silchar-788011,India} \email{atri.deshamukhya@gmail.com}
\affiliation{$^2$ Harish-Chandra Research Institute, Allahabad- 211019, India}\email{panda@mri.ernet.in}

\begin{abstract}
We analyze warm tachyonic inflation, proposed in the literature , but from the viewpoint of four dimensional effective action for tachyon field on a non-BPS D3-brane. We find that consistency with observational data on density perturbation and validity of effective action requires warped compactification. The number of background branes which source the flux is found to be of the order of 10 in contrast to  the order of $10^{14}$ in the standard cold inflationary scenario.

\end{abstract}


\maketitle

\section{\bf Introduction}
The inflationary mechanism \cite{GA,AS} though provides natural
solutions to 'problems' viz. horizon, flatness etc of  Standard
Big Bang model --- can't naturally generate the correct magnitude
of initial density  perturbations demanded from experimental
observations \cite{WMAP}. Standard inflationary scenario is divided
into two regimes --- slow roll expansion and reheating phase which
happen in two different time periods. It is assumed that
exponential expansion places the universe in a super-cooled phase
and subsequently thereafter the universe is reheated. In mid 90s it was
recognized by Berera and Fang \cite{BF} that meshing these two
isolated stages may resolve the disparities created by each
separately. In Ref.{\cite{BF}}, it was shown that slow-roll
inflation \cite{BST} is parametrically consistent with a thermal
component.

The resolution of the horizon problem, which underlies
inflationary cosmology, is that at a very early time, the
equation of state that dictates the expansion rate of the Universe
was dominated by a vacuum energy density  $\rho_v$, so that a
small causally connected patch grew to a size that encompasses the
comoving volume which becomes the observed universe today. In the
standard (isentropic) inflationary scenarios, the radiation energy
density $\rho_\gamma$ becomes negligible  rapidly since it scales
inversely with the fourth power of the scale factor. In such case,
a short time reheating period terminates the inflationary period
initiating the radiation dominated epoch.

 But from general theory of relativity for inflation, $\rho_\gamma$ need not be
 negligible and the condition which needs to be satisfied is  that $\rho_\gamma<\rho_v$.
 Inflation in the presence of non-negligible radiation is thus characterized by a non-isentropic
  expansion \cite{OR} and thermal seeds of density perturbations \cite{BF}. This can be realized in the
  warm inflation scenarios \cite{BB} where there is no need for reheating but it is possible to have a smooth transition to the radiation dominated epoch.

The basic idea of warm inflation is that the inflaton field is
coupled to several other fields. As the inflaton relaxes toward
its minimum energy configuration, it decays into lighter fields,
generating an effective viscosity{\cite{GR}}. If this viscosity is
large enough, the inflaton will reach a slow-roll regime, where
its dynamics becomes over-damped. This overdamped regime has been
analyzed in Ref.\cite{BGR}. As one expects, over-damping is most
successful for the case where the inflaton is coupled to a large
number of fields which are thermally excited, i.e., have small
masses compared to the ambient temperature of radiation. This
result has important consequences for cosmological applications
since overdamping must be efficient to get the correct number of
e-folds before the end of inflation.

 There seems to be other reasons which makes warm inflationary scenario more compelling.
 First, in this scenario, since the macroscopic dynamics of the background field and fluctuations \cite{BF,AB}
 are classical from the onset, there is no quantum-classical transition problem. Thus it overcomes a conceptual
 barrier that the supercooled picture could not get away with. Second, in warm inflation models, in regimes relevant
 to observation, the mass of the inflaton field is typically much larger than the Hubble scale. These models thus do
 not suffer from what is sometimes called the 'eta problem'. Finally, accounting for dissipative effects may be important
 in alleviating the initial condition problem of inflation \cite{BG}.

The inclusion of thermal effects acts similar to a mass term which
breaks the scale-symmetry of the zero-temperature theory
\cite{AB2}. In Ref.\cite{BR} it has been argued that the coupling
constant fine-tuning problem is closely associated to this scale
symmetry and breaking this scale symmetry can avoid this problem. We remark in  passing that there has been debate in the literature \cite{yoko} questioning the very possibility of warm inflation. In this note, we donot attempt to assert on this issue either way. On the other hand, assuming that it is possible,
we argue that just like the cold inflationary scenario \cite{JCDP}, even the warm inflation model needs warp compactification for the validity of low energy effective action obtained from ten 
dimensional string theory.

We organize the paper as follows: in next section we briefly review the tachyon warm inflationary model put forward by Herrera et al \cite{Herrera}.
In section III, we discuss the low energy tachyon effective action and the constraints on parameters coming from compactification. In section IV, we analyze the model including the effect of radiation and discuss the consequences of warm inflation. The last section is devoted to conclusion of our analysis.

\section{  Warm Tachyon Inflation}

In this section we review the model of warm tachyonic inflation proposed by Herrera et al .

 The authors analyzed FRW cosmological model in terms of an effective fluid with energy density, $\rho_T$ and pressure, $p_T$ associated with the tachyon field $T$ and defined respectively by
 
 \begin{equation}
 \rho_T~=~\frac{V(T)}{\sqrt{1 - \dot{T}^2}}, ~~~~~~~~~p_T~=~- V(T) \sqrt{1 - \dot{T}^2}
 \end{equation}
 
 where $V(T)$ is the effective potential associated with the tachyon field. The dynamics of the cosmological model, in the warm inflationary scenario is governed by the equations:
  
 \begin{equation}
H^2~=~\frac{1}{3 M_P^2}(\rho_T + \rho_{\gamma})
\end{equation} ,

\begin{equation}
{\dot{\rho}_T}+3H(\rho_T+p_T)=-\Gamma {\dot{T}}^2 \nonumber
\end{equation}

\begin{equation}
\implies \frac{\ddot{T}}{(1-{\dot{T}}^2)}+3H\dot{T}\left(1+\frac{\Gamma \sqrt{1-{\dot{T}}^2}}{3HV}\right) +\frac{V_{,_T}}{ V}=0
\end{equation}
and
\begin{equation}
{\dot{\rho}}_{\gamma}+4H\rho_{\gamma}=  \Gamma {\dot{T}}^2
\end{equation}
 where $H = \dot{a}/a$ is the Hubble factor, $a$ is the scale factor, $\rho_\gamma$ is the energy density of the radiation field and $\Gamma$ is the dissipation coefficient which is responsible for the decay of the tachyon field into radiation during the inflationary epoch.  The presence of the energy density of radiation and the dissipation coefficient distinguishes the warm inflation from the cold inflation dynamics.
 
 The authors used the potential that corresponds to the tachyon field living on a D-brane  of the bosonic string theory, namely $V(t) = V_0 \exp[{-\alpha T}]$ and derived the constraints on the $V_0$ and $\alpha$ by solving the above set of equations and imposing observational constraints on cosmological parameters coming from the density perturbation theory (see \cite{Herrera} for details).  However, in such an approach the constraints coming from string theory, namely the validity of the low energy effective action (to be discussed below) as well as the height of the tachyon potential, which is fixed for a given brane in terms of its tension, did not  play any role. Thus their analysis is phenomenologically motivated rather than describing a viable inflationary (warm) model which is derived from compactified string theory. On the contrary, these issues have been discussed earlier in the cold inflationary scenario and we briefly review it in the next section.
 
 \section{Tachyon Effective Action and coupling to Gravity:}
 
 The effective action for the tachyon field, living on a non-BPS D3-brane, when coupled to four dimensional gravity can be written as \cite{String} 
 
\begin{equation}\label{action}
S~=~\int d^4x \sqrt{-g}\left(\frac{M_P^2}{2} R~-~A V(T)\sqrt{1+B g^{\mu\nu} \partial_\mu T \partial_\nu T}\right)
\end{equation}
where $V(T)$ is the positive definite tachyon potential which has a maximum at $T = 0$ with normalization $V(0) = 1$ and $V~\rightarrow ~0$ as $T~\rightarrow~\infty$. $A$ and
$B$ are dimensionful constants which depends on string length and the closed string coupling
constant $g$. Note that the tachyon field here is dimensionless.

In the conventional dimensional  reduction approach $A$, $B$ and $M_P^2$ are given by
\begin{eqnarray}
A~&=&\frac{\sqrt{2}}{(2 \pi)^3 g \alpha'^2}\\
B~&=& 8 \ln 2~\alpha' \\
M_P^2~&=&\frac{v}{g^2 \alpha'}
\end{eqnarray}
corresponding to the case of space-filling non-BPS D3-brane in type IIA theory. $v$ is related to the volume $V_6$ of the compact manifold as
\begin{equation}
v~=~\frac{2 V_6}{(2 \pi)^7 \alpha'^3}
\end{equation}
and is a dimensionless constant. Note that the tension of the non-BPS brane, which usually comes with the tachyon potential, has been included in $A$ and we will consider the potential function $V(T)$ to be $\exp[{- T^2}]$. This is motivated from string field theory and gives a good description for small $T$ which is assumed to be accurate for the inflationary epoch.  As mentioned, for the validity of the above effective action one requires $g\ll 1$ and $v \gg 1$.

The energy density ${\rho}_T$ and pressure $p_T$ for
the tachyon field derived from the above action are given as:
\begin{eqnarray}\label{ep}
\rho_T~&=&~\frac{AV(T)}{\sqrt{1-B{\dot{T}}^2}}\\  \nonumber
p_T~&=&~-AV(T)\sqrt{1-B \dot{T}^2}.
\end{eqnarray}

The inflation dynamics is governed by the Einstein equations for FRW background metric and the tachyon field equation derived from the above action \cite{FTCP} (see also \cite{MPP}). These dynamical equations involve the pressure density and energy density as defined in the above and hence depend upon $A$ and $B$. Solving these equations and demanding cosmological parameters to be consistent with observational data put severe constraints on $A$ and $B$.  In particular, it was observed in \cite{KL} that in this formalism, it is difficult to meet the second slow roll condition ($\eta$-problem), besides the problem of ensuring the validity of low energy limit of string compactification; namely   $g \ll 1$  and $v \gg 1$. The solutions to these problems were found in Ref \cite { JCDP} by considering a compactification to produce an warped background where the definition of $A$ and $B$ changes and will be defined in a later section.  It was assumed that all the moduli fields are stabilized and the warped background was created by a stack of D6-branes.  However, to ensure that  the cosmological parameters  to be consistent with the observational  data, it was found that the number of required background D6-branes is $10^{14} $ which seems to be too high. Thus it is  natural to re-examine this issue in the warm inflationary scenario and we carry out this analysis in the next sction. Our interesting observation is that the minimum number of required background branes is less than ten.

\section{Warm inflation in non-warped and warped background:}
 
As mentioned in section II,  in the warm inflationary scenario, i.e when thermal effects are taken into account, the corresponding equations are modified due to the contribution of radiation energy density. For the effective action (\ref{action}) and corresponding energy density and pressure density in (\ref{ep}), the inflation  dynamics is governed by the following equations :
\begin{equation}
H^2~=~\frac{1}{3 M_P^2}(\rho_T + \rho_{\gamma})
\end{equation} ,

\begin{equation}
{\dot{\rho}_T}+3H(\rho_T+p_T)=-\Gamma {\dot{T}}^2 \nonumber
\end{equation}

\begin{equation}
\implies \frac{\ddot{T}}{(1-B{\dot{T}}^2)}+3H\dot{T}\left(1+\frac{\Gamma \sqrt{1-B{\dot{T}}^2}}{3HABV}\right) +\frac{V_{,_T}}{B V}=0
\end{equation}
and
\begin{equation}
{\dot{\rho}}_{\gamma}+4H\rho_{\gamma}=  \Gamma {\dot{T}}^2
\end{equation}
In our notation,  overdots represent
derivative with respect to time and $()_{,_T}\equiv \frac{\partial
}{\partial T}$.

In the inflationary epoch, it is assumed that
\begin{eqnarray}
\rho_T~&\sim&~ AV\\
\rho_T~&>&~\rho_\gamma.
\end{eqnarray}
Since in the slow-roll regime ${\dot{T}}^2\ll 1$ and $\ddot{T}\ll (3H+\frac{\Gamma}{ABV})\dot{T}$ and defining the dissipation rate $r$ as
\begin{equation}
r\equiv\frac{\Gamma}{3HABV}
\end{equation}
the Friedmann equation and the slow-roll equation of motion for the tachyon field respectively takes the form:
\begin{eqnarray}
H^2~&=&~\frac{AV}{3 M_P^2}\\
3HB(1+r){\dot{T}}~&=&~-\frac{V~,_T}{V}.
\end{eqnarray}

For a quasi-stable radiation process during inflationary epoch, we need
\begin{equation}
\dot{\rho_\gamma} \ll 4H\rho_\gamma
\end{equation}
and
\begin{equation}
\dot{\rho_\gamma}\ll \Gamma{\dot{T}}^2.
\end{equation}

Hence from Eq.(10), we obtain
\begin{eqnarray}
\rho_\gamma~&=&~\frac{\Gamma {\dot{T}}^2}{4H}\\
&=&\frac{r~M_P^2}{4 B(1+r)^2}(\frac{V~,_T}{V})^2\nonumber\\
&=&\sigma {T_\gamma}^4
\end{eqnarray}
where $T_\gamma$ is the radiation temperature and $\sigma$ is the
Stephan-Boltzmann constant.

The standard slow-roll parameters can then be written as :
\begin{eqnarray}
\varepsilon&\equiv& -\frac{\dot{H}}{H^2}={\frac{M_P^2}{2 A B (1+r)V}}\left(\frac{V~,_T}{V}\right)^2\\
\eta&\equiv& -{\frac{\ddot{H}}{H\dot{H}}}={\frac{M_P^2}{A B(1+r)
V}}\left[\frac{V~,_{TT}}{V}-{{\frac{1}{2}}\left(\frac{V~,_T}{V}\right)^2}\right].
\end{eqnarray}

End of inflation is marked by $\varepsilon\simeq 1$ which also implies
\begin{equation}
\rho_T\simeq 2\frac{(1+r)}{r}\rho_\gamma.
\end{equation}

Number of e-folds before the end of inflation is given by
\begin{equation}
N(T)~=~-\frac{AB}{M_P^2}\int^{T_e}_T {\frac{V^2}{V~,_T}}(1+r)~dT'.
\end{equation}

Using perturbed FRW metric in the longitudinal gauge and following
the procedure depicted in Ref.\cite{Herrera} one can express the
fluctuation of tachyon field as
\begin{equation}
\delta T~=~ C ( \ln V)~,_T exp(\mathfrak{F}(T))
\end{equation}
where $C$ is a constant and $\mathfrak{F}$ is given by:
\begin{equation}
\mathfrak{F}(T)=~-\int\left[\frac{1}{3H+\Gamma/ABV}\left(\frac{\Gamma}{ABV}\right)_{,_T}+\frac{9}{8}\frac{(2H+\Gamma/ABV)}{(3H+\Gamma/ABV)^2}\frac{(\ln V)_{,_T}}{ABV}\left(\Gamma+4HABV-\frac{\Gamma_{,_T}(\ln V)~,_T}{12HB(3H+\Gamma/ABV)}\right) \right]dT.
\end{equation}

The density perturbation is defined to be \cite{Herrera}
\begin{equation}
\delta_H~=~\frac{2}{5}~\frac{exp[-\mathfrak{F}(T)]}{(\ln V)~,_T}\delta T.
\end{equation}

If one restricts to the region of high dissipation where the
dissipation parameter $\Gamma$ is much higher than the product of
expansion rate of the universe and the potential energy of the
scalar field i.e $\Gamma \gg 3HABV$ then using Eq.(21) in $r\gg
1$ limit one can write
\begin{equation}
\delta_H^2~=~\frac{4}{225}~\frac{exp[-2\tilde{\mathfrak{F}}(T)]}{(H~B~r~\dot{T})^2} (\delta T)^2
\end{equation}
where
\begin{equation}
\tilde{\mathfrak{F}}(T)~=~-\int\left[\frac{1}{3Hr}\left(\frac{\Gamma}{ABV}\right)_{,_T}+\frac{9}{8}(\ln V)_{,_T}\left(1-\frac{\ln \Gamma~,_T(\ln V)_{,_T}}{36H^2~B~r}\right) \right]dT.
\end{equation}

In warm inflationary scenario, the fluctuations of the tachyon field
are generated by thermal interaction with the radiation field and
in the high dissipation regime, following Ref.\cite{TB} this can be
written as:
\begin{equation}
(\delta T)^2\simeq \frac{k_F~T_\gamma}{2\pi^2M_P^2}.
\end{equation}

Here $k_F$ corresponds to the freeze-out scale at which thermally
excited fluctuations are damped by dissipation. This freeze-out wave
number is defined as $k_F~=~H\sqrt{3r}\geq H$.

Using equations (19), (27) and (29) we find
\begin{equation}
\delta_H^2\approx \frac{1}{25\sqrt{3}\pi^2}exp[-2 \mathfrak{F}(T)]\left[\left(\frac{1}{\tilde{\varepsilon}}\right)^3\frac{9}{2r^2\sigma M_P^4 AB^4V}\right]^{1/4}
\end{equation}
where  $\tilde{\varepsilon}$ is the slow-roll parameter in the $r\gg 1$ limit.

Note that in the high dissipation regime the slow-roll parameters given by equations (20) and (21) take the forms--
\begin{eqnarray}
\tilde{\varepsilon}~&=~&\frac{M_P^2}{2 rABV}\left[\frac{V~,_T}{V}\right]^2\\
\tilde{\eta}~&=&~\frac{M_P^2}{ rABV}\left[\frac{V~,_{TT}}{V}-\frac{1}{2}\left(\frac{V~,_T}{V}\right)^2\right]
\end{eqnarray}

The scalar spectral index is given by:
\begin{eqnarray}
n_s-1~&=&~\frac{d\ln \delta_H^2}{d\ln k} \nonumber \\
&=&~\frac{3\tilde{\eta}}{2}+\tilde{\varepsilon}\left[\frac{2V}{V~,_T}\left(2\tilde{\mathfrak{F}}(T),_T-\frac{r~,_T}{4r}\right)-\frac{5}{2}\right]
\end{eqnarray}
and hence the running of the spectral index is
\begin{eqnarray}
&\alpha_s&~=~\frac{dn_s}{d\ln k}\nonumber\\
&=&-\frac{2V\tilde{\epsilon}}{V,_T}\left\{\frac{3\tilde{\eta},_T}{2}
+\frac{\tilde{\epsilon},_T}{\tilde{\epsilon}}\left(n_s-1-\frac{3\tilde{\eta},_T}{2}\right)
+2\tilde{\epsilon}\left[\left(\frac{V}{V,_T}\right),_T\left(2\tilde{\mathfrak{F}},_T-\frac{(\ln r),_T}{4}\right)+\left(\frac{V}{V,_T}\right)\left(2\tilde{\mathfrak{F}},_{TT}-\frac{(\ln r),_{TT}}{4}\right)
\right]
\right\}
\end{eqnarray}

Power spectrum of tensor modes in this scenario has an extra
temperature dependence factor $\coth\left[\frac{k}{2T}\right]$(see
Ref.\cite{KSA}) and is given by:
\begin{equation}
A_g^2~=~\frac{H^2}{2~\pi^2~M_P^2} \coth \left[\frac{k}{2T_\gamma}\right]
\end{equation}
Thus the tensor spectral index is
\begin{eqnarray}
n_T~&=&~\frac{d}{d\ln k} \ln\left[\frac{A_g^2}{\coth[k/2T_\gamma]}\right]\nn
&=&-2\varepsilon.
\end{eqnarray}
Power spectrum of the scalar modes on the other hand is given by
\begin{eqnarray}
P_\mathcal{R}~&=&~\frac{25}{4}\delta_H^2\nonumber\\
&=&\frac{1}{2\sqrt{2}}~\left[\frac{A^2}{M_P^{10}~B}~r~\frac{V^8}{V,_T^6}\right]^{1/4}\exp[-2\tilde{\mathfrak{F}}].
\end{eqnarray}
Hence tensor to scalar ratio takes the form
\begin{eqnarray}
R(k_0)~&=&~\frac{A_g^2}{P_\mathcal{R}}\Large{|}_{k=k_0}\nonumber\\
&=&\frac{\sqrt{2}\sigma^2}{3}~\left[\frac{A^2~B}{M_P^6~r}~\frac{V^{12}}{V_{,T}^6}\right]^{1/4}\exp[2\tilde{\mathfrak{F}}]~\coth \left[\frac{k}{2T_\gamma}\right]\vert_{k=k_0}.
\end{eqnarray}

\subsection{Numerical Analysis}
As motivated earlier, we now work with the explicit potential function $V(T)~=~e^{-T^2}$.
The dissipation coefficient $\Gamma$ can de related to the radiation temperature using Eq.(19).And we have
\begin{equation}
\tilde{\mathfrak{F}}(T)~=~-2\frac{V''}{V'}\left(1-\frac{3}{8}\frac{\sigma T_\gamma^4}{AV}\right)+\frac{3}{8}\frac{V'}{V}\left(1-\frac{\sigma T_\gamma^4}{2A}\frac{V'}{V}\right).
\end{equation}

Similarly, using Eq.(19), one can also express the slow roll parameters in terms of radiation temperature as follows:
\begin{eqnarray}
\tilde{\varepsilon}~&=&~2\sigma~\frac{T_\gamma^4}{A}~\frac{1}{V}\\
\tilde{\eta}~&=&~4\sigma ~\frac{T_\gamma^4}{A}~\left[\frac{V,_{TT}}{V_T^2}-\frac{1}{2V}\right].
\end{eqnarray}

In terms of these parameters, the cosmological observable can be expressed as
\begin{eqnarray}
n_s&=&1+\frac{3}{2}~\tilde{\eta}+~\tilde{\varepsilon}~\left[-\frac{3 \sigma T_\gamma^4}{4 A}\frac{V'}{V}-\frac{3(\sigma T_\gamma^4/A-3V) V~,_{TT}}{V~,_T^2}\right]\\
n_T&=&-\tilde{\varepsilon}\\
\alpha_s&=&-\frac{2V\tilde{\varepsilon}}{V,_T}
\left[\frac{3\tilde{\eta},_T}{2}+\tilde{\varepsilon},_T\left(-\frac{3 \sigma T_\gamma^4}{4 A}\frac{V'}{V}-\frac{3(\sigma T_\gamma^4/A-3V) V~,_{TT}}{V~,_T^2}\right)\right] \nonumber \\ 
&-&\frac{4V\tilde{\varepsilon}^2}{V,_T}\left[\left(\frac{V}{V,_T}\right),_T\left(2\tilde{\mathfrak{F}},_T-\frac{(\ln r),_T}{4}\right)+\frac{V}{V,_T}\left(2 \tilde{\mathfrak{F}},_{TT}-\frac{(\ln r),_{TT}}{4}\right)\right]\\
P_\mathcal{R}
&=&\frac{1}{4\pi^2~\sigma^2}~~\left[\frac{A^2}{M_P^8~B^2~T_\gamma^4}~r~\frac{V^6}{V,_T^4}\right]^{1/4}\exp[-2\tilde{\mathfrak{F}}]\\
R(k_0)
&=&\frac{2\sigma^2}{3}~\left[T_\gamma^4~\frac{A^2~B^2}{M_P^8}~\frac{V,_T^4}{V^2}\right]^{1/4}\exp[2\tilde{\mathfrak{F}}]~\coth \left[\frac{k}{2T_\gamma}\right]\Large{|}_{k=k_0}.
\end{eqnarray}

All these observable quantities are to be evaluated at $T_*$, which is the value of the tachyon field at roughly 60 e-folds before the end of inflation.  We consider the following strategy for our numerical estimation of $A$ and $B$ which will lead to the analysis of constraints on $g$ and $v$:

1. From the condition for end of inflation,
$\tilde{\varepsilon}(T_e)~=~1$, $T_e$ is obtained as a function of
${T_{\gamma}^4}/{A}$.

2. Fixing $N_e$ to be 60, $T_*$ is evaluated as a function of $T_e$.

3. The lower limit for $T_\gamma^4/A$ is  found out for which
the upper bound on running of spectral index is consistent with the observational data whereas the upper limit is fixed from the condition that minimum  value of $N_e$ is 60.

4. Using WMAP data \cite{WMAP} for $P_\mathcal{R}$ i.e.  $P_\mathcal{R}~\leq~2.3\times 10^{-9}$, range of $A^2/(T_\gamma^4M_P^8B^2)$ is obtained.

We quote below the range for $A/M_P^4$, $\sqrt{A/M_P^{6}B}$ obtained from our above numerical estimation.
\begin{eqnarray}
10^{-9}\leq \frac{A}{M_P^4}< 10^{-3} \\
2.5\leq \sqrt{\frac{A} {M_P^{6}B}}\leq 10^{32}
\end{eqnarray}
In this analysis we have used $k_0~=~0.002Mpc^{-1}$ and $T_\gamma=0.24\times10^{16}GeV$ \cite{Herrera}. $\sigma$ is taken to be of the order of unity. We have confirmed that the slow roll conditions are satisfied for these values of $A$ and $T_\gamma$.

Using the above we predict the bounds on the rest of the observable as below :
The spectral index is in the range
\begin{equation}
0.9480\leq n_s\leq 0.9636.
\end{equation}
Running of the spectral index is found to be negative and lie in the range
\begin{equation}
-0.00093\leq \alpha_s\leq -0.00099.
\end{equation}
The tensor-scalar ratio is found to be negligible over the whole range of parameters. These are well within the limit given by WMAP 5years data \cite{WMAP}.

We now proceed to analyse the above results in terms of the string theory parameters $g$ and $v$. For each corresponding set of values of $A/M_P^4$ and $A^2/M_P^{12}~B^2$ we can solve for $g$ and $v$ as :
\begin{eqnarray}
g^7~&=&~\frac{2^{13/2}~\ln 2}{(2 \pi)^3}~\frac{A^2/M_P^{12}~B^2}{(A/M_P^4)^3}\\
v^{14}~&=&~\frac{2^{10} \ln 2}{(2 \pi)^24}\frac{A2/M_p^{12}B^2}{(A/M_P^4)^{10}}.
\end{eqnarray}

We tabulate below the values of $g$ and $v$ corresponding to the range of  $A/M_P^4$ and $A^2/(M_P^{12}~B^2)$ obtained earlier.
\vskip 1cm
\begin{table}[!htb]
\begin{center}
\begin{tabular}{|c|c|c|c|}
\hline
  $A/M_P^4$ & $\sqrt{\frac{A}{M_P^{6}B}}$ & $g$ & $v$ \\
\hline
  $10^{-9}$ & $2.48$ & $9.9\times10^{3}$ & $2.38\times10^{5}$ \\
\hline
$10^{-7}$ & $7.85\times10^{10}$ & $1.38\times10^{9}$ & $8.87\times10^{6}$ \\
\hline
$10^{-5}$ & $4.28\times10^{21} $& $1.38\times10^{21}$ & $3.86\times10^{8}$ \\
\hline
$10^{-3}$ & $7.84\times10^{31} $& $2.67\times10^{19}$ & $1.23\times10^{10}$ \\
\hline
\end{tabular}
\vskip .5cm
Table I
\end{center}
\end{table}

From this table it is clear that in the allowed
parameter range for $A$ and $B$ for which the cosmological observables are consistent with experimental data, the required constraint  $g\ll 1$ and $v\gg1$ is never satisfied simultaneously.

\subsection{Warm Inflation in Warped background}

As concluded in the previous section, within conventional compactification it is not possible to obtain physical parameters of inflation consistent with observations keeping $g\ll 1$ and $v\gg 1$. To solve this problem, we now redo our analysis considering the warped compactification \cite{warp} similar to the analysis of \cite{JCDP} for the case of cold inflation.

We introduce warping  by considering the ten-dimensional string frame metric of the form
\begin{equation}
ds^2~=~e^{2C(y)}g_{\mu\nu}(x)dx^{\mu}dx^{\nu}+g_{mn}(y){dy^m}dy^n
\end{equation}
where $e^{2C(y)}$ being the warp factor which can take very small
values. Here $x$ and $y$ denote the coordinates on the four-dimensional non-compact and sis-dimensional compact space respectively. In addition, the dilaton field is allowed to vary over the
compact manifold as
\begin{equation}
\phi~=~{\phi_0}+\phi(y).
\end{equation}
Such warping can be produced by introducing a number of background D6-branes and an appropriate number of O6-planes for charge conservation.

In this background, clearly the definitions of $A$ and $B$ gets modified as:
\begin{eqnarray}
A~&=&~\frac{\sqrt{2}e^{4C-\phi}}{{(2 \pi)^3} g {\alpha'}^2}\\
B~&=&~8\ln{2\alpha' e^{-2C}}.
\end{eqnarray}
The functions $C$ and $\phi$ here are subject to the solutions of
equations of motion derived from supergravity theory.

The four dimensional Planck mass in such a compactification is found to be
\begin{equation}
{M_P}^2~=~\frac{\tilde{v}}{g^2\alpha'}
\end{equation}
where $g=e^{\phi_0}$ and the warped volume $\tilde{v}$ is
\begin{equation}
\tilde{v}={\frac{2}{(2\pi)^7 {\alpha'}^3}} \int{d^6 y
\sqrt{g_6}e^{-2\phi+2C}}.
\end{equation}
By choosing average value of $e^{-2\phi+2C}$ of the order of one, $\tilde{v}$ can be taken of same order as $v$.

Within the supergravity approximation one can derive the following expressions for the warp factor and the dilaton field:
\begin{eqnarray}
e^{2C}&=&(g N_{min})^{-2/3}\\
e^{-\phi}&=&~g~N_{min}
\end{eqnarray}
where $N_{min}$ is the minimum number of background D6-branes. For slow-roll conditions to be satisfied one requires $e^{4C-\phi} \gg 1$.

In such a background, the relation of $g$ and $v$ with parameters $A$ and $B$ depends on the warp factors and are given by:
\begin{eqnarray}
g~e^{\phi}~&=&~\left(\frac{2^{13/2}~\ln 2}{(2 \pi)^3}~\frac{A^2/M_P^{12}~B^2}{(A/M_P^4)^3}\right)^{\frac{1}{7}}\\
v~e^{-2C+\phi}~&=&~\left(\frac{2^{10} \ln 2}{(2 \pi)^{24}}\frac{A2/M_p^{12}B^2}{(A/M_P^4)^{10}}\right)^{\frac{1}{14}}
\end{eqnarray}

Eliminating $e^\phi$ and $e^{2C}$ by the help of equations (62) and (63), the relation between $g$ and $v$ is found to be:
\begin{equation}
g~=~2563~\frac{(A/M_P^4)^{12/7}}{(A^2/M_P^{12}B^2)^{1/14}}~v^3
\end{equation}

Following our analysis in the previous section, we furnish the reinterpreted values of string coupling $g$ and volume of compact space $v$ in the light of warped background in table below.
\vskip 1cm
\begin{table}[!htb]
\begin{center}
\begin{tabular}{|c|c|c|c|c|c}
\hline
  $A/M_P^4$ & $\sqrt{A/M_P^{6} B}$ & $g e^{\phi}$ & $v e^{-2C+\phi}$& $v$ for$g=.01$ \\
\hline
  $10^{-9}$ & $2.48$ & $9.9\times10^{3}$ & $2.38\times10^{5}$ & $2363$\\
\hline
$10^{-7}$ & $7.85\times10^{10}$ & $1.38\times10^{9}$ & $8.87\times10^{6}$ & $1716$\\
\hline
$10^{-5}$ & $4.28\times10^{21} $& $1.38\times10^{21}$ & $3.86\times10^{8}$ & $1301$\\
\hline
$10^{-3}$ & $7.84\times10^{31} $& $2.67\times10^{19}$ & $1.23\times10^{10}$ & $889$ \\
\hline
\end{tabular}
\vskip .5cm
Table II
\end{center}
\end{table}
We observe that it is possible to achieve, simultaneously, $g\ll 1$ and $v\gg 1$ for the whole range of papameters.

Using the equations (62) and (63) and the condition coming from slow-roll i.e $e^{4C-\phi}\gg 1$, the minimum number of background D6- branes required to produce the required background is found to be less than ten. This is a significant improvement over the cold inflation model where the minimum number of such branes required was found to be $10^{14}$ \cite{JCDP}.

\section{discussion}

In this work we investigated the viability of warm tachyonic inflationary scenario . Our analysis revealed that, though with a string theory motivated inflaton potential it is possible to get all the cosmological observables within the experimental bounds, it is not possible to comply with the validity of low energy effective action obtained from  the conventional toroidal compactification namely the string coupling constant to be  much less than unity and the volume of compact space to be larger than unity. On the contrary these problems could  be resolved in an warped compactification. Moreover, an interesting observation is that  the number of background D6-branes required to produce the necessary warped background turned out to be less than ten. This can be contrasted with  the corresponding number $10^{14}$  in a cold inflation scenario. Our analysis assumed that all moduli fields are stabilized. We point out that it has been observed in \cite{HK} that it is impossible, in general flux compactification scheme, to cure the "$\eta$-problem" of inflation models derived from Type IIA string theory (the case we studied here) where any of the moduli fields play the role of an inflaton. Thus, in such a setup, it may be a welcome fact to trade the tachyon field as the inflaton.  In principle, the introduction of a non-BPS brane can have some effect, through a change in K$\ddot{a}$hler potential, on the moduli stabilization, but we have not taken this into account in our present analysis. We hope to report on this in future. However, in the context of cold inflation, this issue is addressed partially in \cite{JW}.

\section*{Acknowledgements}
AD would like to thank the members, HRI, Allahabad for their continuous support to visit  the Institute  during which most of the work has been carried out.



\begin{thebibliography}{99}
\bibitem{GA}
  A.~H.~Guth,
  Phys.\ Rev.\  D {\bf 23} (1981) 347-352.
 \bibitem{AS}
  A.~J.~Albrecht and P.~J.~Steinhardt,
  Phys.\ Rev.\ Lett.\  {\bf 48}, 1220-1223 (1982).
  A.Linde, Particle Physics and inflationary cosmology (Gordon and Breach, New York, 1990).
\bibitem{WMAP}
  E.~Komatsu {\it et al.}  [WMAP Collaboration],
  Astrophys.\ J.\ Suppl.\  {\bf 180}, 330-376(2009)
  [arXiv:0803.0547 [astro-ph]].


\bibitem {BF} A.~Berera and L.~Z.~Fang,
  Phys.\ Rev.\ Lett.\  {\bf 74}, 1912-1915 (1995)
  [arXiv:astro-ph/9501024].


\bibitem{BST}  J.~M.~Bardeen, P.~J.~Steinhardt and M.~S.~Turner,
  Phys.\ Rev.\  D {\bf 28}, 679 (1983).
   \\
    A.~D.~Linde,
  Phys.\ Lett.\  B {\bf 129}, 177-181 (1983).



\bibitem{OR} H.~P.~de Oliveira and R.~O.~Ramos,
  Phys.\ Rev.\  D {\bf 57}, 741-749 (1998)
  [arXiv:gr-qc/9710093].


   E.~Gunzig, R.~Maartens and A.~V.~Nesteruk,
  Class.\ Quant.\ Grav.\  {\bf 15}, 923-932 (1998)
  [arXiv:astro-ph/9703137].



  A.~Berera,
  Phys.\ Rev.\  D {\bf 55}, 3346-3357 (1997)
  [arXiv:hep-ph/9612239].


\bibitem{BB} A.~Berera,
             Phys.\ Rev.\ Lett.\  {\bf 75}, 3218 (1995)[arXiv:astro-ph/9509049].

             A.~Berera,
             Phys.\ Rev.\  D {\bf 54}, 2519-2534 (1996)
             [arXiv:hep-th/9601134].



\bibitem{GR} M.~Gleiser and R.~O.~Ramos,
  Phys.\ Rev.\  D {\bf 50}, 2441-2455 (1994)
  [arXiv:hep-ph/9311278].
\\
             A.~Hosoya and M.~a.~Sakagami,
  Phys.\ Rev.\  D {\bf 29}, 2228 (1984).
    \\
             M.~Morikawa,
  Phys.\ Rev.\  D {\bf 33}, 3607 (1986).



\bibitem{BGR} A.~Berera, M.~Gleiser and R.~O.~Ramos,
  Phys.\ Rev.\  D {\bf 58}, 123508 (1998)
  [arXiv:hep-ph/9803394].

\bibitem{AB} A.~Berera,
  Nucl.\ Phys.\  B {\bf 585}, 666 -714(2000)
  [arXiv:hep-ph/9904409].


\bibitem{BG}
  A.~Berera and C.~Gordon,
  Phys.\ Rev.\  D {\bf 63}, 063505 (2001)
  [arXiv:hep-ph/0010280].

  R.~O.~Ramos,
  Phys.\ Rev.\  D {\bf 64}, 123510 (2001)
  [arXiv:astro-ph/0104379].



\bibitem{AB2} A.~Berera,
  Phys.\ Rev.\ Lett.\  {\bf 75}, 3218 (1995)
  [arXiv:astro-ph/9509049].



\bibitem{BR} B.~Ratra,
  Phys.\ Lett.\  B {\bf 260}, 21-26 (1991).
  
  \bibitem{yoko}
  J.~Yokoyama and A.~D.~Linde,
  Phys.\ Rev.\  D {\bf 60}, 083509 (1999)
  [arXiv:hep-ph/9809409].

  
  \bibitem{JCDP} J.~Raeymaekers,
  JHEP {\bf 0410}, 057 (2004)
  [arXiv:hep-th/0406195].



               P.~Chingangbam, S.~Panda and A.~Deshamukhya,
               JHEP {\bf 0502}, 052 (2005)
               [arXiv:hep-th/0411210].


\bibitem{Herrera} R.~Herrera, S.~del Campo and C.~Campuzano,
  JCAP {\bf 0610}, 009 (2006)
  [arXiv:astro-ph/0610339].



\bibitem{String} A.~Sen,
  JHEP {\bf 9910}, 008 (1999)
  [arXiv:hep-th/9909062].


               M.~R.~Garousi,
  Nucl.\ Phys.\  B {\bf 584}, 284-299 (2000)
  [arXiv:hep-th/0003122].


               E.~A.~Bergshoeff, M.~de Roo, T.~C.~de Wit, E.~Eyras and S.~Panda,
               JHEP {\bf 0005}, 009 (2000)[arXiv:hep-th/0003221].


               J.~Kluson,
              Phys.\ Rev.\  D {\bf 62}, 126003 (2000)
              [arXiv:hep-th/0004106].



\bibitem{FTCP} M.~Fairbairn and M.~H.~G.~Tytgat,
              Phys.\ Lett.\  B {\bf 546}, 1-7 (2002)
              [arXiv:hep-th/0204070].


               D.~Choudhury, D.~Ghoshal, D.~P.~Jatkar and S.~Panda,
               Phys.\ Lett.\  B {\bf 544}, 231 (2002)
               [arXiv:hep-th/0204204].



\bibitem{MPP} A.~Mazumdar, S.~Panda and A.~Perez-Lorenzana,
             Nucl.\ Phys.\  B {\bf 614}, 101 (2001)
              [arXiv:hep-ph/0107058].



\bibitem{KL}  L.~Kofman and A.~Linde,
  JHEP {\bf 0207}, 004 (2002)
  [arXiv:hep-th/0205121].






\bibitem{TB} A.~N.~Taylor and A.~Berera,
  Phys.\ Rev.\  D {\bf 62}, 083517 (2000)
  [arXiv:astro-ph/0006077].


\bibitem{KSA} K.~Bhattacharya, S.~Mohanty and A.~Nautiyal,
  Phys.\ Rev.\ Lett.\  {\bf 97}, 251301 (2006)
  [arXiv:astro-ph/0607049].



\bibitem{warp} 

  C.~S.~Chan, P.~L.~Paul and H.~L.~Verlinde,
  Nucl.\ Phys.\  B {\bf 581}, 156 (2000)
  [arXiv:hep-th/0003236].




  I.~R.~Klebanov and M.~J.~Strassler,
  JHEP {\bf 0008}, 052 (2000)
  [arXiv:hep-th/0007191].




  S.~B.~Giddings, S.~Kachru and J.~Polchinski,
  Phys.\ Rev.\  D {\bf 66}, 106006 (2002)
  [arXiv:hep-th/0105097].



  B.~R.~Greene, K.~Schalm and G.~Shiu,
  Nucl.\ Phys.\  B {\bf 584}, 480-508 (2000)
  [arXiv:hep-th/0004103].


\bibitem{HK} M.~P.~Hertzberg, S.~Kachru, W.~Taylor and M.~Tegmark,
  JHEP {\bf 0712}, 095 (2007)
  [arXiv:0711.2512 [hep-th]].



\bibitem{JW} J.~Ward,
  arXiv:0901.0080 [gr-qc].


\end{thebibliography}
\end{document}